\documentclass[doublecol,final]{epl2}
\usepackage[T1]{fontenc}
\usepackage[latin9]{inputenc}
\usepackage{graphicx}

\makeatletter
 
 \@ifundefined{textcolor}{}
 {%
   \definecolor{BLACK}{gray}{0}
   \definecolor{WHITE}{gray}{1}
   \definecolor{RED}{rgb}{1,0,0}
   \definecolor{GREEN}{rgb}{0,1,0}
   \definecolor{BLUE}{rgb}{0,0,1}
   \definecolor{CYAN}{cmyk}{1,0,0,0}
   \definecolor{MAGENTA}{cmyk}{0,1,0,0}
   \definecolor{YELLOW}{cmyk}{0,0,1,0}
 }

\usepackage{bm}
\usepackage{graphics}
\usepackage{booktabs}
\PassOptionsToPackage{caption=false}{subfig}

\makeatother

\begin{document}

\title{Intersections of moving fractal sets}

\author{I. Mandre \and J. Kalda}

\institute{Institute of Cybernetics at the Tallinn University of Technology,
Akadeemia tee 21, 12618, Tallinn, Estonia}
\abstract{
Intersection of a random fractal or self-affine set with a linear
manifold or another fractal set is studied, assuming that one of the
sets is in a translational motion with respect to the other. It is
shown that the mass of such an intersection is a self-affine function
of the relative position of the two sets. The corresponding Hurst
exponent $h$ is a function of the scaling exponents of the intersecting
sets. A generic expression for $h$ is provided, and its proof is
offered for two cases --- intersection of a self-affine curve with
a line, and of two fractal sets. The analytical results are tested
using Monte-Carlo simulations.
}




\maketitle
There is a wide spectrum of problems which can be reduced to finding
and studying intersections of fractal sets. For instance, rain intensity
is a multifractal function of space and time \cite{Lovejoy1985,Lovejoy1990};
rainfall at a given point on Earth's surface is a time-integral of
this function --- a measure of the intersection of the rain intensity
field with a line, parallel to the time axis. Next, the Doppler absorption
spectra depend on how many points of the flow move with a certain
velocity in a certain direction; in the case of fully turbulent flows,
velocity is a self-affine function of coordinates: the problem is
reduced to finding the number of intersection points of a self-affine
curve and a straight line. Further, it has been shown that silicate
clay deposits, when drying, collapse into a self-affine surface; in
the case of fractal deposits, the dry surface height is defined by
the size of the intersection of the fractal with a vertical line \cite{Fossum2004}.

While the list of examples could be further extended, we stop after
providing just one more example which is discussed in more detail:
reflection of light from surfaces with non-smooth gradients. More
specifically, we consider surfaces with fractional Brownian (fB) gradient
components; an example of such surfaces is provided by a free water
surface in the case of fully developed wave turbulence \cite{Zakharov1967,Dyachenko2004}.

Suppose a collimated beam of light falls onto a two-dimensional surface
described by its height $z=z\!\left(x,y\right)$, such that the gradient
components $\partial_{x}z=f\!\left(x,y\right)$ and $\partial_{y}z=g\!\left(x,y\right)$
are fB functions, so that
\begin{equation}
\left\langle \left[f\!\left(x,y\right)\!-\! f\!\left(x',y'\right)\right]^{2}\right\rangle \propto\left[\left(x-x'\right)^{2}+\left(y-y'\right)^{2}\right]^{H},
\end{equation}
and the same scaling law holds for $g\!\left(x,y\right)$. Here, angular
braces denote averaging over an ensemble of surfaces, and $H$ is
the Hurst exponent, $0<H<1$. Let us assume that the lower cut-off
scale of this scaling law is unity, and that the gradient components
become smooth below that scale. Furthermore, we assume that the wavelength
of the incident light is much smaller than one. The functions $f$
and $g$ define a two-dimensional random self-affine surface (the
gradient surface) $u=f\!\left(x,y\right),$ $v=g\!\left(x,y\right)$
in four-dimensional space $x,y,u,v$. The propagation direction of
incident light from a point on the surface $z$ is determined by its
gradient components at that point. Therefore, the intensity of light
reflected at a given direction from the entire surface $z$ is proportional
to the number of intersection points $N$ of the gradient surface
with a two-dimensional linear manifold $u=u_{0},v=v_{0}$. This number
is a random function of the propagation direction, $N=N\!\left(u,v\right)$
--- the light intensity fluctuates as the observation direction is
changed. We will show that the function $N\!\left(u,v\right)$ can
be described by another Hurst exponent $h=h\!\left(H\right):$
\begin{equation}
\left\langle \left[N\!\left(u,v\right)\!-\! N\!\left(u',v'\right)\right]^{2}\right\rangle \propto\left[\left(u-u'\right)^{2}+\left(v-v'\right)^{2}\right]^{h}.\label{eq:scaling-law}
\end{equation}

We start by deriving the dependance $h=h\left(H\right)$ for the case
of a fB curve intersecting with a line in two-dimensional space. Let
us consider a finite-length segment $x\in\left[0,L\right]$ of a fB
curve $u=f\!\left(x\right)$ with zero mean. Then, typically, the
curve varies from $u\sim-L^{H}$ to $u\sim L^{H}$ (here ``$\sim$''
means ``is of the order of''). The fractal dimension of the fB curve
is $2-H$ \cite{Orey1970}, and the dimension of its intersection
with a line $u=u_{0}$ is $d_{f}=1-H$ \cite{Mandelbrot,Mandelbrot1984}.
The intersection at level $u=0$ is also known as the zero set.

Since the lower cut-off scale is unity, the number of intersection
points at some fixed $u_{0}$ is estimated as $N\!\left(u_{0}\right)\sim L^{d_{f}}=L^{1-H}.$
Let us denote the change in the number of intersection points when
changing from ``altitude'' $u_{0}$ to $u_{0}+\Delta u$ as $\Delta N\!\left(\Delta u\right)\equiv N\!\left(u_{0}\right)-N\!\left(u_{0}+\Delta u\right).$
We now make use of a scale-decomposition of the function $f\!\left(x\right)$
by introducing course-grained functions $F_{a}\!\left(x\right)=\sum_{2^{i}\geq a}f_{2^{i}}\left(x\right)$,
where $f_{2^{i}}\left(x\right)$ are the scale components that can
be obtained, for instance, via a forward and reverse Fourier transform
of $f\left(x\right)$, where only the wavelengths between $2^{i-1}$
and $2^{i}$ are kept. Let us denote the number of intersection points
of the line $u=u_{0}$ with the course-grained curve $u=F_{a}\!\left(x\right)$
as $N_{a}\!\left(u_{0}\right)\sim\left(L/a\right)^{1-H}$ and the
change in the intersection points due to displacement $\Delta u$
as $\Delta N_{a}\!\left(\Delta u\right)\equiv N_{a}\!\left(u_{0}\right)-N_{a}\!\left(u_{0}+\Delta u\right).$
As we increase the level difference $\Delta u$, the line $u=u_{0}+\Delta u$
will cross from time to time the extrema of the function $u=F_{a}\!\left(x\right)$.
By each crossing, the number of intersections $N_{a}\!\left(u_{0}+\Delta u\right)$
will change by two --- increase in the case of a minimum, and decrease
in the case of a maximum. When $\Delta u\ll a^{H}$, that is it is
well below the vertical characteristic scale, these changes are purely
incidental --- they are caused by uncorrelated extrema that are separated
by large distances. Therefore, at $\Delta u\ll a^{H}$, the value
$\Delta N_{a}\!\left(\Delta u\right)$ is a compound Poisson process,
that is $\Delta N_{a}\!\left(\Delta u\right)=\sum_{i=1}^{P\!\left(\Delta u\right)}D_{i}$,
where $\left\{ P\!\left(\Delta u\right):\Delta u\geq0\right\} $ is
a Poisson process with rate $\lambda$, and $\left\{ D_{i}:i\geq1\right\} $
are independent random values drawn with equal probability from $\left\{ -2,+2\right\} $.
The variance of the compound Poisson process \cite{Tijms2003} is
$\lambda\Delta u\left\langle D^{2}\right\rangle $; but as $\Delta N_{a}\!\left(\Delta u\right)$
has zero mean, we conclude that
\begin{equation}
\left\langle \Delta N_{a}\!\left(\Delta u\right)^{2}\right\rangle =4\lambda\Delta u\quad\left(\Delta u\ll a^{H}\right).\label{eq:course-grained-delta-N_a-at-small-displa}
\end{equation}
We estimate the density of extrema for $N_{a}$ as $\lambda\sim\left(L/a\right)/L^{H}=L^{1-H}a^{-1}$
--- the number of peaks is $L/a$ and they are distributed quasi-homogeneously
in the range $-L^{H}$ to $L^{H}$. We note that Eq. (\ref{eq:course-grained-delta-N_a-at-small-displa})
can also be used to estimate the scaling exponent for displacements
below the lower cut-off scale of $f\!\left(x\right)$, that is for
$\Delta u\ll1$, we obtain a super-universality $h\!\left(H\right)=1/2$.

Around each intersection point with the course-grained curve, the
line $u=u_{0}$ also intersects with the fine-scaled structure $f\!\left(x\right)-F_{a}\!\left(x\right)$.
But as the intersection points with the course-grained curve are typically
spaced at greater distances than $a$, the number of intersections
with the fine-scaled structure around each such point are uncorrelated
(as correlations in the fine-scaled structure only extend to distances
around $a$). We denote the average number of such intersections around
each point as $n_{a}\sim a^{1-H}$ and conclude that the total number
of intersections with the whole curve $u=f\!\left(x\right)$ is $N\!\left(u_{0}\right)\sim n_{a}N_{a}\!\left(u_{0}\right)\sim a^{1-H}N_{a}\!\left(u_{0}\right)$.
As we move the intersecting line from level $u_{0}$ to $u_{0}+\Delta u$,
the number of intersections with $u=f\!\left(x\right)$ changes. When
the displacement $\Delta u$ is smaller than $a^{H}$, the contributions
from the fine-scaled intersections are highly correlated, but at displacement
$\Delta u\gg a^{H}$ they are basically uncorrelated. Consequently,
\begin{equation}
\Delta N\!\left(\Delta u\right)\sim a^{1-H}\Delta N_{a}\!\left(\Delta u\right)\quad\left(\Delta u\gg a^{H}\right).\label{eq:delta-N-at-big-displacement}
\end{equation}
At the marginally applicable limit $\Delta u=a^{H}$, the equations
(\ref{eq:course-grained-delta-N_a-at-small-displa}) and (\ref{eq:delta-N-at-big-displacement})
combine into
\begin{equation}
\left|\Delta N\!\left(a^{H}\right)\right|\sim a^{1-H}\left|\Delta N_{a}\!\left(a^{H}\right)\right|\sim L^{\frac{1-H}{2}}a^{\frac{1-H}{2}}.
\end{equation}
To estimate $\left|\Delta N\!\left(\Delta u\right)\right|$, we choose
$a=\Delta u^{1/H}$, yielding 
\begin{equation}
\left|\Delta N\!\left(\Delta u\right)\right|\sim L^{\frac{1-H}{2}}\Delta u^{\frac{1-H}{2H}},
\end{equation}
and so the scaling exponent $h$ for the intersection of a fB curve
and a moving line is
\begin{equation}
h=\frac{1-H}{2H}\quad\left(1\ll\Delta u\ll L^{H}\right).\label{eq:1d-h}
\end{equation}

It should be noted that for $H<\frac{1}{3}$, this equation yields
$h>1$. Result $h>1$ means that large-scale fluctuations are so strong
that the gradients of large-scale components dominate over the gradients
caused by small-scale fluctuations. In that case, Eq. (\ref{eq:scaling-law})
would yield the Hurst exponent $h=1$. However, using wavelet or Fourier
analysis, it is possible to generalize Eq. (\ref{eq:scaling-law})
and reveal scaling laws with $h>1$.

Returning to the case of the intensity of light reflected by the sea
surface, where we have an intersection of a self-affine surface and
a flat surface in 4D space, the scaling law (\ref{eq:scaling-law})
can be derived in a similar fashion, resulting in $h=\frac{2-2H}{2H}$,
where $2-2H$ is the fractal dimension of the intersection studied.

\begin{figure}
\includegraphics[width=1\columnwidth]{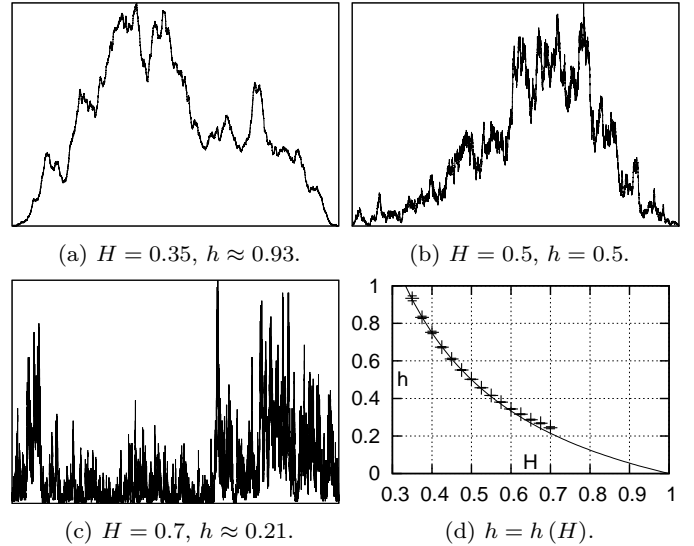}

\caption{\label{fig:samples}(a)-(c): sample functions of $N\!\left(u\right)$
for various $H$; (d): Monte-Carlo results for the scaling of the
intersection of a fractional Brownian curve and a moving line, solid
line is the predicted value.}
\end{figure}

We have run a series of Monte-Carlo simulations to test the result
(\ref{eq:1d-h}). At each calculation point $H$ we generated $1000$
fractional Brownian curves $f\left(x\right)$ with length $L=2^{27}$
\cite{DAVIES1987,Dietrich1997,Wood1994}. Samples of the intersection
functions $N\!\left(u\right)$ can be seen in Fig.~\ref{fig:samples}.
The data was analyzed using the continuous wavelet transform and the
Mexican hat wavelet \cite{Simonsen1998}. The results follow the predicted
relationship $h=\frac{1-H}{2H}$ quite closely except at greater values
of $H$ {[}Fig.~\ref{fig:samples}(d){]}. The discrepancy is due
to distortions in the function $N\!\left(u\right)$ --- as $L^{H}$
grows, the density of intersections falls and the function $N\!\left(u\right)$
starts to experience large ranges where it is of constant small value
{[}see Fig.~\ref{fig:samples}(c) for a sample $N\!\left(u\right)$
at $H=0.7${]}. This is a finite-size effect --- to overcome it one
would have to calculate at much greater length $L$. We also did some
calculations for $H<1/3$. The results were as expected with $h>1$.

We now turn our attention to general statistically self-similar fractal
sets. It is easy to imagine that the interactions (changes in the
intersection) of a line and a random fractal set at displacements
well below the lower scaling length of the fractal are completely
random. We have also found, that for the intersections of a fB curve
or surface with a line or a plane, the analytically derived scaling
exponents all came out as $h=d_{f}/\left(2H\right)$, where $d_{f}$
is the fractal dimension of the intersection. Considering all this,
one can conjecture that this relation also applies to intersections
of general random fractals. We proceed to make this claim more specific.

Let us have two fractal sets $\mathcal{F}$ and $\mathcal{X}$ with
corresponding fractal dimensions $d_{\mathcal{F}}$ and $d_{\mathcal{X}}$.
Let the set $\mathcal{X}$ be translatable in some direction $\hat{\mathbf{u}}$,
with the position identified by coordinate $u$. Further, we assume
that it is self-similar and with finite scaling range $\left[1,L\right]$.
This set may also have a topological dimension that is equal to its
fractal dimension (for example, it may be a simple line or a plane).
We assume that the fractal set $\mathcal{F}$ is random, that is it
is only statistically self-similar (we will clarify the nature of
this randomness further along the way). Let the fractal $\mathcal{F}$
have the same scaling range as $\mathcal{X}$ in the directions perpendicular
to $\hat{\mathbf{u}}$ but let it be possibly self-affine in the direction
$\hat{\mathbf{u}}$ with scaling range $\left[1,L^{H_{\hat{\mathbf{u}}}}\right]$.

The fractal dimension of the intersection of the two sets is $d_{f}=d_{\mathcal{F}}+d_{\mathcal{X}}-D,$
where $D$ is the dimension of the surrounding space. As the set $\mathcal{X}$
moves, the total fractal mass of this intersection $M\!\left(u\right)$
(the number of points, the surface area, the volume, or other such
measure that is suitable for the given fractal depending on its topological
dimension) will change. We fix $u=u_{0}$ and denote this change at
translation to $u=u_{0}+\Delta u$ as $\Delta M\!\left(\Delta u\right)\equiv M\!\left(u_{0}\right)-M\!\left(u_{0}+\Delta u\right).$
We conjecture that the function $M\!\left(u\right)$ is fractional-Brownian-motion-like,
that is it can be described by
\begin{equation}
\left\langle \Delta M\!\left(\Delta u\right)^{2}\right\rangle \propto\left|\Delta u\right|^{2h},\label{eq:general-fractal-scaling-law}
\end{equation}
with the Hurst exponent $h$ as
\begin{equation}
h=\frac{d_{f}}{2H_{\hat{\mathbf{u}}}},\label{eq:master-conjectured-relation}
\end{equation}
where $H_{\hat{\mathbf{u}}}$ describes the scaling of the fractal
$\mathcal{F}$ in the direction $\hat{\mathbf{u}}$ ($H_{\hat{\mathbf{u}}}$
is unity for a self-similar fractal set).

We will now continue with a derivation leading to this result for
the case of self-similar fractals (with $H_{\hat{\mathbf{u}}}=1$).
For this we will first approximate the fractals by the use of a ball
cover --- this results in a ``course-grained'' version of the fractal
at a specific grain size. Then, we will derive how $M\!\left(u\right)$
scales at movements either much smaller or much greater than the length
used at the ball cover. Finally, we bring these two estimates together
to yield the exponent $h$.

In case the set $\mathcal{F}$ is self-similar, the fractal mass of
the intersection $\mathcal{F}\cap\mathcal{X}$ can be estimated as
$M\!\left(u_{0}\right)\sim L^{d_{f}}.$ Let us assume that we can
find minimal covers for both sets $\mathcal{F}$ and $\mathcal{X}$
with $D$-dimensional closed balls of diameter $a$, where $1\ll a\ll L$.
The number of balls in either cover can be estimated as $N_{\mathcal{F}}\!\left(a\right)\sim\left(L/a\right)^{d_{\mathcal{F}}}$
and $N_{\mathcal{X}}\!\left(a\right)\sim\left(L/a\right)^{d_{\mathcal{X}}}$.

At a location where two balls, each from a different set, intersect,
the fractal sets themselves usually intersect, with the average fractal
mass of the intersection (assuming $\mathcal{F}$ is random, for example
the balls can't be globally aligned) estimated as $m_{a}\sim a^{d_{f}}$.
The total number of such intersections is $N_{a}\!\left(u_{0}\right)\sim M\!\left(u_{0}\right)/m_{a}\sim\left(L/a\right)^{d_{f}}$.
We move the set $\mathcal{X}$ in the direction $\hat{\mathbf{u}}$
by distance $\Delta u$. This will cause the cover of the set $\mathcal{X}$
also move. As a ball from that cover moves, it penetrates or exits
balls covering the standing set $\mathcal{F}$. As a result, the value
$N_{a}\!\left(u_{0}\right)$ will increase or decrease by one. We
denote the total change in the number of intersecting balls as $\Delta N_{a}\!\left(\Delta u\right)\equiv N_{a}\!\left(u_{0}\right)-N_{a}\!\left(u_{0}+\Delta u\right)$.

In case the movement is much greater than $a$, that is $a\ll\Delta u\ll L$,
a moving ball that is penetrating a standing ball will exit it. We
assume that $\mathcal{F}$ is random in such a way that the masses
of the sub-fractals contained in individual standing balls separated
by distances much greater than $a$ are uncorrelated. In such a case
\begin{equation}
\Delta M\!\left(\Delta u\right)\sim m_{a}\Delta N_{a}\!\left(\Delta u\right)\quad\left(a\ll\Delta u\ll L\right).\label{eq:gen-scal-mass-at-big-displacement}
\end{equation}

In case the movement is much smaller than $a$, that is $\Delta u\ll a$,
a moving ball that is intersecting a standing ball will rarely exit
it. Also, it has very little chance to interact with other standing
balls or the correlations in their placement (defined by the structure
of the fractal). With small movement individual moving balls have
no chance to interact with the fractal structure of the ball cover.
And assuming the fractal $\mathcal{F}$ is random, that is the balls
are not globally aligned, we can ignore their interactions as a group.
In such a case the change in the number of balls intersected can be
approximated as a compound Poisson process, that is $\Delta N_{a}\!\left(\Delta u\right)=\sum_{i=1}^{P\!\left(\Delta u\right)}D_{i}$,
where $\left\{ P\!\left(\Delta u\right):\Delta u\geq0\right\} $ is
a Poisson process with rate $\lambda$, and $\left\{ D_{i}:i\geq1\right\} $
are independent random values drawn with equal probability from $\left\{ -1,+1\right\} $.
The variance of the compound Poisson process is $\lambda\Delta u\left\langle D^{2}\right\rangle $;
but as $\Delta N_{a}\!\left(\Delta u\right)$ has zero mean, we conclude
that
\begin{equation}
\left\langle \Delta N_{a}\!\left(\Delta u\right)^{2}\right\rangle =\lambda\Delta u\quad\left(\Delta u\ll a\right).\label{eq:gen-scaling-of-N_a-at-small-displacement}
\end{equation}

We now estimate the Poisson process rate $\lambda.$ At displacement
$\Delta u$ the moving balls cover the volume $V_{\mathcal{X}}\sim N_{\mathcal{X}}\!\left(a\right)a^{D-1}\Delta u\sim L^{d_{\mathcal{X}}}a^{D-1-d_{\mathcal{X}}}\Delta u$.
Assuming the standing balls are distributed quasi-homogeneously (the
fractal $\mathcal{F}$ is random), their density per volume of space
is $\rho_{\mathcal{F}}\sim N_{\mathcal{F}}\!\left(a\right)/L^{D}\sim L^{d_{\mathcal{F}}-D}a^{-d_{\mathcal{F}}}.$
The number of balls encountered during movement $\Delta u$ must then
be $N_{\Delta u}\sim\rho_{\mathcal{F}}V_{\mathcal{X}}\sim L^{d_{f}}a^{-1-d_{f}}\Delta u$.
The rate of balls encountered is $\lambda\sim N_{\Delta u}/\Delta u\sim L^{d_{f}}a^{-1-d_{f}}$.

At the marginally applicable limit $\Delta u=a$ of the two expressions
(\ref{eq:gen-scal-mass-at-big-displacement}) and (\ref{eq:gen-scaling-of-N_a-at-small-displacement}),
we estimate the change in the mass as
\begin{equation}
\left|\Delta M\!\left(a\right)\right|\sim m_{a}\sqrt{\lambda}a^{1/2}\sim L^{d_{f}/2}a^{d_{f}/2}.
\end{equation}
Since the ball cover size $a$ can be freely chosen between $1$ and
$L$, we can pick $a=\Delta u$, confirming conjecture (\ref{eq:general-fractal-scaling-law})
with the Hurst exponent $h=d_{f}/2.$

For the case $H_{\hat{\mathbf{u}}}\neq1$, one would have to take
into account that the correlations in the self-similar structure of
the fractal scale at a different rate in the direction $\hat{\mathbf{u}}$.

The intersection of two fractals may have a dimension less than $0$.
Previously, it has been interpreted as how ``empty'' the intersection
is \cite{Mandelbrot1984,Mandelbrot1990}. In equation (\ref{eq:master-conjectured-relation})
this would result in negative $h$. This is not necessarily a pathological
case, as a negative $h$ can be used when instead of (\ref{eq:general-fractal-scaling-law})
the scaling is given through the Fourier power spectrum, that is through
the relation $\left\langle \left|\psi_{\mathbf{k}}\right|^{2}\right\rangle \propto\left|\mathbf{k}\right|^{-2h-1}$.
However, we have not tested this numerically.

\begin{figure}
\includegraphics[width=1\columnwidth]{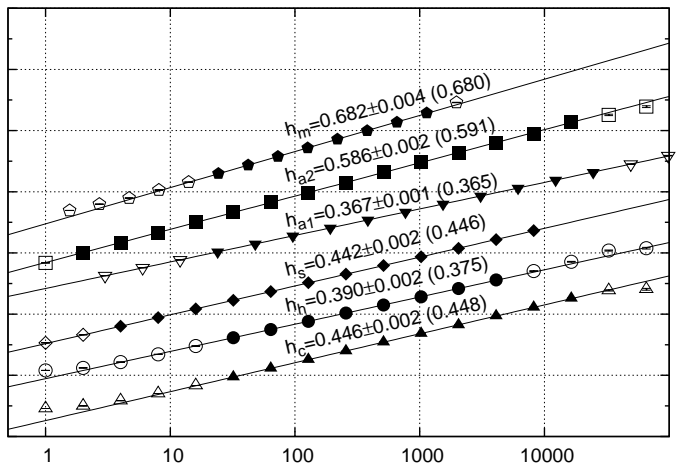}\caption{\label{fig:all-results-on-graph}Wavelet based scaling exponent fitting
for the intersections of random percolation cluster ($h_{c}$) and
hull ($h_{h}$); the randomized Sierpi\'{n}ski carpet ($h_{s}$);
the self-affine randomized Sierpi\'{n}ski carpet ($h_{a1}$ and $h_{a2}$);
and the intersection of a percolation cluster with a deterministic
Sierpi\'{n}ski carpet ($h_{m}$). Results for different fractals have
been moved up or down to fit on the same graph. Uncertainties are
given with $0.95$ significance. The predicted values $h=d_{f}/\left(2H\right)$
are in the parentheses. Only filled points were used for the fits.}
\end{figure}

To test the relation (\ref{eq:master-conjectured-relation}) we ran
Monte-Carlo simulations for the following cases: two-dimensional random
bond percolation cluster and hull intersected with a horizontal line,
with predicted $h_{c}=\left(91/48+1-2\right)/2$ and $h_{h}=\left(7/4+1-2\right)/2$
for the cluster and hull respectively; randomized $3\times3$ Sierpi\'{n}ski
carpet \cite{Sierp1918,McMullen1984,Bedford1984,Peres2008,Gui2008}
(with one cell cleared randomly at each construction step) intersected
with a horizontal line, with predicted $h_{s}=(\log_{3}8+1-2)/2$;
self-affine randomized $4\times3$ Sierpi\'{n}ski carpet (with one
cell cleared randomly at each construction step) intersected with
vertical and horizontal lines, with the carpet's box dimension $d_{4\times3}=\log_{3}\left(3^{1-\log_{4}3}11^{\log_{4}3}\right)$
and predicted scaling exponents $h_{a1}=\frac{\log_{3}11/4}{2\log_{3}4}$
and $h_{a2}=\frac{d_{4\times3}+1-2}{2\log_{4}3}$; percolation cluster
intersected with a deterministic $3\times3$ Sierpi\'{n}ski carpet,
with predicted $h_{m}=\left(91/48+\log_{3}5-2\right)/2$. The results
from the Monte-Carlo simulations are all very close to the predicted
values (Fig.~\ref{fig:all-results-on-graph}). As we increased calculation
lattice sizes we saw improvement across the board, indicating that
the small discrepancies are due to the finite size effects.

To conclude, it is now easy to see that the flow rate of the river
Nile, famously studied by Harold Edwin Hurst \cite{Hurst1951}, is
an integral quantity of the fractal structure of precipitation \cite{Lovejoy1985,Lovejoy1990}
over its drainage basin, and as confirmed by the analytical relation
we have found, is self-affine. This analytical relation should be
applicable in both predictive and descriptive capacity for many problems,
from the matter distribution of the universe to the formation of $1/f$-like
noise in semiconductor devices.
\acknowledgments
This work was supported by Estonian Science Targeted Project No. SF0140077s08,
and EU Regional Development Fund Centre of Excellence TK124. The authors
would like to thank the Department of Computer Engineering at the
Tallinn University of Technology for providing computational resources.
\bibliographystyle{eplbib}
\bibliography{intersect}

\end{document}